\documentclass[10pt,a4paper]{article}
\usepackage{amsmath}
\usepackage{amsthm}
\numberwithin{equation}{section}
\usepackage{graphicx,lscape}
\title{\Large \bf
Evidence of Economic Regularities and Disparities of Italian Regions
From Aggregated Tax Income Size Data}
\author{Roy Cerqueti$^{1,}$\footnote{Corresponding address: University of Macerata, Department of Economics and Law,   via Crescimbeni 20, I-62100, Macerata, Italy. Tel.: +39 0733 258 3246; Fax: +39 0733 258 3205. Email: roy.cerqueti@unimc.it} $\;$ and  Marcel Ausloos$^{2,3}$}

 \date{
$^1$ University of Macerata, Department of Economics and Law,\\  via
Crescimbeni 20,   I-62100, Macerata, Italy \\  $e$-$mail$ $address$:
roy.cerqueti@unimc.it\\  \vskip0.5cm $^{2}$eHumanities
group\footnote{Associate Researcher}$\;$, \\Royal Netherlands
Academy of Arts and Sciences, \\  Joan Muyskenweg 25, 1096 CJ
Amsterdam, The Netherlands \\ \vskip0.5cm $^3$GRAPES\footnote{Group
of Researchers for Applications of Physics in Economy and Sociology
Universit\' e de Liege, Sart Tilman,  B-4000 Liege, Belgium}$\;$,
\\   rue de la Belle Jardiniere 483, \\B-4031, Angleur, Belgium \\$e$-$mail$ $address$:
marcel.ausloos@ulg.ac.be
 }
\begin{document}
 \maketitle

\begin{abstract}

This paper discusses the size distribution, - in economic terms - of
the Italian municipalities over the period 2007-2011. Yearly data
are rather well fitted by a modified Lavalette law, while
Zipf-Mandelbrot-Pareto law seems to fail in this doing. The analysis
is performed either at a national as well as at a local (regional
and provincial) level. Deviations are discussed as originating in so
called king and vice-roy effects. Results confirm that Italy is
shared among very different regional realities. The case of Lazio is
puzzling.

\end{abstract}

\section{ Introduction}\label{Introduction}

The analysis of the ranking of elements belonging to a specific set
under a predefined criterion leads to the identification of a best
fit\footnote{All fits, in this communication,  are based on the
Levenberg-Marquardt  algorithm (Levenberg 1944, Marquardt  1963,
Lourakis  2011); the error bar was pre-imposed to  be at most 1\%.}
curve, through the rank-size theory  (Jefferson 1939, Zipf 1949,
Beckmann 1958, Gabaix 1999a, Gabaix 1999b) and its applications.
\newline This paper deals with the rank-size rule for the entire set
of municipalities in Italy (IT, hereafter) for each year of the
quinquennium 2007-2011. The \textit{size} is here given by the
\textit{contribution} (so called Aggregated Tax Income, thereby
denoted hereafter as ATI) that each city has given to the Italian
GDP (data are expressed in Euros);  cities are yearly ranked
according to the value of their related ATI. Data are official, and
have been provided directly from the Research Center of the Italian
Minister of Economic Affairs.
\newline For our
investigation, several different directions are followed:
\begin{itemize}
\item[$1.$] the possible law describing the relationship
between ranking and ATI is explored. In particular, we show that
Zipf, Zipf-Mandlebrot\footnote{It is sometimes called the
Zipf-Mandlebrot-Pareto  (ZMP) function.} and power laws fail in this
doing. A more convincing answer is provided by  the Lavalette
function (Popescu, 2003),
 \begin{equation} \label{Lavalette2}
y(rank)=K\;  \Big(\frac{N\cdot rank}{N-rank+1}\Big)^{-\chi}\;\equiv\; \kappa\;  \Big(\frac{rank}{N-rank+1}\Big)^{-\chi}
\end{equation}
which has been introduced in 1996 by the biophysicist Daniel
Lavalette. Such an analysis is performed not only at the country,  but
also at the regional and at the provincial level;
\item[$2.$] the distribution of the ATI at the regional level
is lengthily explored. In doing so, several cities are shown to
exhibit a prominent role in determining a relevant percentage of the
national GDP (the so-called \textit{king} and \textit{king plus
vice-roy effects}, see Section \ref{ATIregions} for the details).
\end{itemize}
In particular, point 1. supports that sometimes data  city sizes do
not have pure Zipf-type (i.e. a pure power law)  links with the
corresponding ranks. However, evidence is here shown that some
particular subsets of cities may be well described by a
statistically appealing  Zipf-Mandelbrot  law (this is the
paradigmatic case of Lazio, an IT region), - a set of considerations
postponed for an Appendix (App. A) in order to let   a relatively
ordered line of thought guiding the reader in the following
sections, - without  being distracted by the main aims. For the contextualization of these results in the
literature, see Section \ref{Review}. \newline Also point 2. is in
great agreement with an improvement of the best-fit results when
some specific subsets of data are considered. In this case, king and
king plus vice-roy effects can be appreciated by observing, on
displayed plots, that removing the first and sometimes the first set
of ranked cities, respectively, leads (not always, but remarkably
often) to a more statistically convincing Lavalette curve.
\newline It is important to point out that, to the best of our
knowledge, this is the first contribution dealing with the
application of the Lavalette curve to the field of urban economics;
it was  invented  and usually applied for bibliometrics studies.
\newline The paper is organized as follows: Section \ref{Review}
briefly reviews the literature inspiring and connected to the
present research.  Section \ref{8092} contains the description of
the data. Section \ref{citydistributions} is devoted to the
investigation of the whole IT, with the assessment of some rank-size
rule fits on yearly basis. This section contains also the ATI
ranking analysis at a regional level, with  all the plots of the
2-parameter Lavalette functions and the detection of the outliers.
Section \ref{sec:results} collects and discusses the findings. The
last section  (Sect. \ref{conclusions}) concludes and offer
suggestions for further research lines. Appendix A describes the
Lazio case, while Figures and Tables pertaining to the regional data
analysis are collected in Appendix B.

\begin{figure}
\includegraphics[height=6.0cm,width=9.8cm] 
{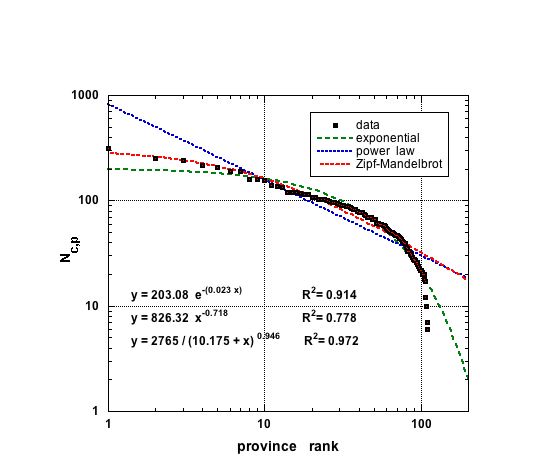}
 \caption   {Log-log plot  of the number $N_{c,p}$ of
cities (8092) per  provinces (110), ranked by decreasing order of
"importance"; showing fits by a power law, an exponential and a
Zipf-Mandlebrot  function with the corresponding correlation
coefficient.} \label{fig:Plotnogood9ZMPpwldexp}
\end{figure}

       \begin{figure}
\includegraphics[height=6.0cm,width=9.8cm] 
{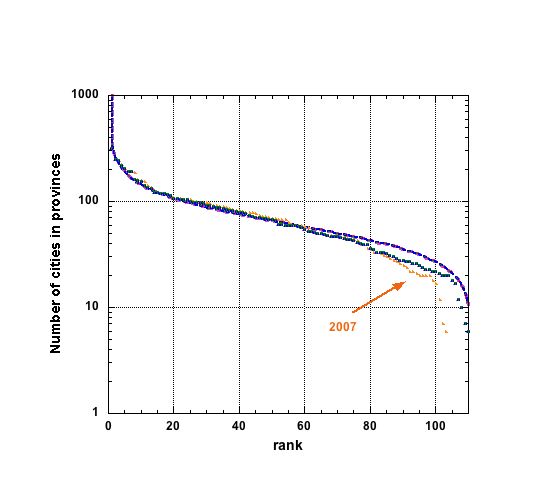}
 \caption   {Semi-log- plot of the number, $N_{c,p}$, of
cities in a province ranked by decreasing order of "importance", for
the  studied 5
 years (2007-2011);  
the best mere 2-parameter Lavalette function  fit, Eq.
(\ref{Lavalette2}), is   shown for year 2011 only for better
visibility ($R^2$=0.985); the 2007 year, with 5 less provinces, is
also emphasized; all best  2-parameter Lavalette  function  fits are
found in Table \ref{TableprovLavfitt}.}
 \label{fig:Plot2Ncplilof}
\end{figure}

  \begin{figure}
\includegraphics  [height=6.0cm,width=9.8cm] 
{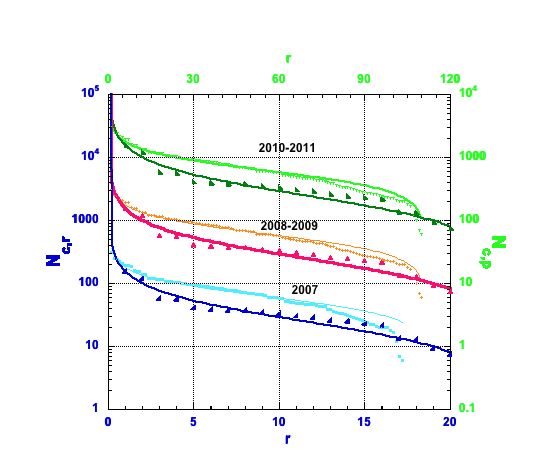}
 \caption   {Semi-log- plot of the number of cities  in a
province, $N_{c,p}$,  and in a region, $N_{c,r}$,  ranked by
decreasing order of "importance", for various years;   the  2007,
2008-2009 and 2010-2011 data are displaced by an obvious factor for
better readability; the best mere 2-parameter Lavalette function,
Eq. (\ref{Lavalette2}),  fit is   shown for  $N_{c,p}$, with
$N_{p}$ =110;  a forced 2-parameter Lavalette function with $N_{r}=
24$  (instead of $N_{r}= 20$) is used for $N_{c,r}$ to improve
$R^2$; all best  2-parameter  Lavalette function  fits are found in
Table   \ref{TableprovLavfitt}. }
 \label{fig:Plot4liloNcp3fNcr3f}
\end{figure}

\section{Review of the literature}\label{Review}
In the context of New Economic Geography (NEG), - introduced by
Krugman (1991) and surveyed in Ottaviano and Puga (1998), Fujita et
al. (1999), Neary (2001), Baldwin et al. (2003) and Fujita and Mori
(2005),  spatial patterns based on geographical agglomerations and
dispersions of economic quantities play a fundamental role. In
discussing the features of the geographical entities, city
population size distribution represents one of the most debated
themes, and there is a wide literature discussing on how the
rank-size rule can be properly described.
\newline
In this respect, power law and Pareto distribution with coefficient
one (the so-called  Zipf's law, introduced in Zipf (1935, 1949),
stating  that a hyperbolic relationship exists between rank and
size), seems to provide a rather satisfactory answer. Several
studies proved empirically the validity of Zipf's law: Rosen and
Resnick (1980) analyzed data from 44 Countries, and found a clear
predominance of statistical significance of Zipf's law, with $R^2$
greater than 0.9 (except in one case, Thailand); in Mills and
Hamilton (1994), data from US cities in 1990 has been taken to show
the evidence of Zipf's law ($R^2\sim 0.99$); other papers which
substantially support this type of rank-size rule are Guerin-Pace
(1995), Dobkins and Ioannides (2001), Song and Zhang (2002),
Ioannides and Overman (2003), Gabaix and Ioannides (2004), Reed
(2002), Dimitrova and Ausloos (2013) just to cite a few. Nitsch (2005)
provides an exhaustive literature review up to that time. It is also
worth mentioning Simon (1955), Gabaix (1999a, 1999b) and Brakman et
al. (1999), who have the merit to have tried to provide an
explanation of Zipf's law. However, Gabaix (1999b) criticized Simon
(1955) reasoning in saying that it is grounded on assumptions on the
Pareto parameter that seem to be not empirically supported.
\newline Recently,
Dimitrova and Ausloos (2013), through the notion of the global
primacy index of Sheppard  (1985) indicated that Gibrat  (growth)
law (Gibrat, 1931), supposedly at the origin of Zipf's law, in fact,
does not hold in the case of Bulgaria cities.
\newline
Thus, in general, why the rank-size rule can be described in many cases
through the Zipf's law remains still a puzzle. This lack of a
theoretical basis for this statistical results has been acknowledged
by influential scientists (see Fujita et al., 1999; Fujita and
Thisse, 2000).
\newline
Moreover, Zipf's law is not a  universal law at all, in the sense
that some data does not support such a way to link rank and size of
the cities. As an example, the above-mentioned case of Thailand in
Rosen and Resnick (1980) concerns a weak correlation between data
plot and Pareto fit. Peng (2010) found a Pareto coefficient of
$0.84$ -not so close to one!, - when implementing a best fit of data
on Chinese city sizes in 1999-2004 through Pareto distribution.
Ioannides and Skouras (2013), like others, argue that Pareto law
seems to stand in force only in the tail of the data distribution.
Matlaba et al. (2013) provided evidence that, at least for the
analyzed case of Brazilian urban areas over a spectacularly wide
period (1907-2008), Zipf's law is clearly rejected.
\newline
The failure of Zipf's law may depend often on the way data are
grouped (Giesen and S\"{udekun}, 2011). In this respect, Soo (2007)
proves empirically that the size of Malaysian cities cannot be
plotted according to such rank-size rule, but a suitable collection
of them can do it. A list of other contributions on the
inconsistency of Zipf's law in several countries, different periods
and under specific economic conditions should include Cordoba
(2008), Garmestani et al. (2007) and Bosker et al. (2008). Of
particular interest is also Garmestani et al. (2008), who conduct an
analysis for the US at a regional level.
\newline
From the present state of the art point of view, regional
agglomerations,  commonly ranked in terms of population, may be also
sorted out in an order dealing with the economic variables. In fact,
Zipf's law is sometimes  identified also in some ''economic'' way to
rank. As an example, Skipper (2011) used such  a rank-size
relationship to  detect   well developed countries order  through
their national GDP. This result has been also achieved by Cristelli
et al. (2012), who exhibited evidence of the Zipf's law for the top
fifty richest countries in the period 1900-2008. One can then
conclude as McCann (2013) does, in stating that \textit{[Zipf's law
holds] irrespective of whether the regional size is measured in
terms of population or GDP.} This is in contrast with Nobel laureate
Krugman previous statement that the rank-size rule is \textit{"a
major embarrassment for economic theory: one of the strongest
statistical relationships we know, lacking any clear basis in
theory."} (Krugman, 1995, p.44).
\newline
No need to say that, therefore, more data analysis can bring some
information on resolving the controversy. Moreover, the
investigation seems new, since there is, to our knowledge,  no
statistical evidence of Zipf's law studies for the economic
variables characterizing Italian cities (in the period 2007-2011).
\newline
Note that  investigations of the contributions (= sizes) that local
entities  bring to the national GDP  have been often studied. Those
investigations are the main themes of  so many publications that
references cannot  be even short listed. However, much literature
has been  rather concerned with convergence effects (as in L\'{
o}pez-Bazo et al., 1999) which have not been the main themes of the
present investigation. Rather than  searching for effects, we have
been aiming at observing and   quantifying structural causes.

\section{Data}\label{8092}
Data collect the disaggregated contributions at a municipal level
(in IT a \textit{municipality} or \textit{city} is denoted as
\textit{comune}, - plural $comuni$) to the Italian GDP.
\newline
The data source is the Research Center of the Italian Minister of
Economic Affairs, and the covered period is the quinquennium
2007-2011.
\newline Under an administrative point of view, Italy
is composed of 20 regions, more than 100 provinces and more than
8000 municipalities\footnote{For a more detailed explanation of the
regional areas, in the framework of EU, refer to the Eurostat at:
$http://epp.eurostat.ec.europa.eu/portal/page/portal/nuts_nomenclature/introduction$.}.
Each municipality is included in one specific province, which in
turns belongs to one and only one region. Several administrative
laws modified the number of provinces and municipalities during the
quinquennium, and also of the number of cities in each entity, but
the number of regions has been constantly equal to 20 (see below the
time dependence of the precise values).
\newline Therefore, the available yearly ATI data corresponds to a
different number of cities. In particular, the number of cities has
been yearly evolving respectively  as follows : 8101, 8094, 8094,
8092, 8092,  -  from 2007 till 2011. \newline However, scientific
consistency imposes to compare identical lists. In 2011, the number
of provinces and municipalities is 110 and 8092, respectively. We
have considered this latest 2011 "count" as the basic one.
Therefore, we have taken into account a virtual merging of cities,
in the appropriate (previous to 2011) years, according to IT
administrative law statements (see also
$http://www.comuni-italiani.it/regioni.html$).
\newline In brief, several cities have thus merged into new ones,
other were phagocytized. Here below are the various cases "of
interest" explaining  some "data reorganization":

\begin{itemize}
\item[(i)]
Campolongo al Torre (UD) and Tapogliano (UD) have merged after a
public consultation, held on Novembre 27th, 2007, into Campolongo
Tapogliano (UD); thus 2 $\rightarrow$ 1
\item[(ii)] LEDRO (TN)
was the result of the merging (after a public consultation, held on
Novembre 30th, 2008) of
 Bezzecca (TN), Concei (TN), Molina di Ledro (TN), Pieve di Ledro (TN),
Tiarno di Sopra (TN) and Tiarno di Sotto (TN) as far as it is
explained  e.g. in
$http://www.tuttitalia.it/trentino-alto-adige/18-concei/$; thus 6
$\rightarrow$ 1
\item[(iii)] Comano Terme (TN) results from the merging of
Bleggio Inferiore (TN) and Lomaso (TN), in force of a regional law
of November 13th, 2009; thus 2 $\rightarrow$ 1
\item[(iv)] Consiglio di Rumo (CO) and Germasino (CO) were annexed by
Gravedona (CO) on May 16th, 2011 and February 10th, 2011, to form
the new municipality of Gravedona ed Uniti (CO); thus 3
$\rightarrow$ 1.
\end{itemize}
To sum up:  13 $\rightarrow$ 4.
\newline
Thus, 8092 municipalities is our reference number. In short, the ATI
(studied in Sect. \ref{citydistributions} and in Sect.
\ref{regionaldisparities}) of the resulting cities have been
linearly adapted, as if these were preexisting before the merging or
phagocytosis. A summary of the statistical characteristics for the
year-dependent ATI of all Italian cities over the period 2007-2011
can be found in Table \ref{Tablestat}. Table
\ref{topandbottomcitiesATIranked} contains the yearly ranked top and
bottom cities in Italy in the sample period.

\begin{table} \begin{center}
\begin{tabular}[t]{ccccccc}
  \hline
   $ $   &2007 &2008 &2009&2010&   2011 & \\
\hline
min. (x$10^{-5}$)   &3.0455         &2.9914      &   3.0909    &3.6083        &3.3479&   \\
Max. (x$10^{-10}$)&   4.3590&4.4360 &    4.4777      &4.5413 &4.5490 \\
Sum (x$10^{-11}$)&6.8947 &7.0427 &    7.0600 &7.1426 &7.2184      \\
mean ($\mu$) (x$10^{-7}$)   &8.5204  &8.7033     &   8.7248&8.8267 &8.9204  \\
median ($m$) (x$10^{-7}$)  &2.2875 &2.3553 &   2.3777 &2.4055&2.4601  \\
RMS (x$10^{-8}$)    &6.5629 &6.6598 &    6.6640&6.7531 &6.7701 \\
Std. Dev. ($\sigma$) (x$10^{-8}$) &6.5078&6.6031&   6.6070& 6.6956 &6.7115 \\
Var.    (x$10^{-17}$)&4.2351&4.3601&    4.3653 &4.4831 &4.5044 \\
Std. Err. (x$10^{-6}$)&7.2344 &7.3404  &   7.3448&7.4432 &7.4609 \\
Skewness  &48.685 &48.855&   49.266&49.414 &49.490 \\
Kurtosis      &2898.7     &2920.42   &   2978.1       &2991.0 &2994.7      \\  \hline
 $\mu/\sigma$  &0.1309   &0.1318&0.1321 &0.1319&  0.1329&  \\
$3(\mu-m)/\sigma$  &0.2873&0.2884&0.2883 &0.2878& 0.2889&   \\
\hline
\end{tabular}
   \caption{Summary of  (rounded) statistical
characteristics  for ATI of IT cities ($N=8092$) in
2007-2011.}\label{Tablestat}
\end{center} \end{table}

Note that, in this time window, the data  claims a number
of 103 provinces in 2007, with an increase by 7 units (BT, CI, FM,
MB, OG, OT, VS) thereafter, leading to 110 provinces. In this
respect, it is worth noting a discrepancy between what data say and
the real legislative evolution of the provinces. In fact, 4
provinces have been instituted by a regional law of 12 July 2001 in
Sardinia and became operative in 2005 (CI, MB, OG, OT), while BT, FM
and VS have been created on June 11th, 2004 and became operative on
June 2009. However, the official data provided by the Economics
Minister are here taken as scientific basis, and the number of
provinces is then 103, 110, 110, 110, 110 - from 2007 till 2011.
\newline
Some (mild) effect of this $provincial$ variation is discussed
below, although the emphasis of the present discussion is about the
$regional$ level.

\section{Regional and provincial analysis}
\label{citydistributions} \vskip 0.5truecm  In order to stress the
regional aspect, the number of cities per regions, and  also per
provinces, ranked in decreasing order of "importance" is examined,
i.e. the number of cities in  a region or  in a  province is the
"size measure", in this section; see Figs.
\ref{fig:Plotnogood9ZMPpwldexp}-\ref{fig:Plot4liloNcp3fNcr3f}:

\begin{itemize}
\item on Fig. \ref{fig:Plotnogood9ZMPpwldexp} it is seen that a mere
2-parameter decaying power law (blue) or a 2-parameter decaying
exponential (green) as well as a 3-parameter Zipf-Mandelbrot
function (red) are  neither visually nor statistically appealing
(see the $R^2$ value) for describing the number of cities in the
provinces as function of the rank, $N_{c,p}(rank)$. Therefore,
further  specific investigations are needed to assess the data.
These are however beyond the scope of the present paper, limiting
ourselves here to fits based on only a 2-parameter function;

\item in contrast,  Fig. \ref{fig:Plot2Ncplilof}, a double $x$ - double $y$ plot,
reports a  fit of the ranking of the 110 provinces, according to the
number of cities, $N_{c,p}$ by a 2-parameter Lavalette function, Eq.
(\ref{Lavalette2}). It seems to be a rather  good fit, to say the
least, with $R^2=0.985$. Some deviation occurs at high rank
($r\ge60$), but there are not many cities (less than 50) in each of
these  few provinces. The 5 yearly cases are hardly distinguishable
from each other. Observe some  different data range for 2007: recall
that there are 7 provinces less in 2007 than in other subsequent
years. To better distinguish the various years, Fig.
\ref{fig:Plot4liloNcp3fNcr3f} shows the rank size variation for
$N_{c,p}$, the number of cities in each province,  fitted with the
appropriate 2-parameter Lavalette function.
\end{itemize}

\begin{table} \begin{center}
\begin{tabular}[t]{cccccc}
  \hline
   $ $   &2007 &2008 &2009&2010&   2011 \\
\hline
 $N_c$   &8101      &8094       &   8094    &8092       &8092    \\
 $N_p$  &103    &110    &110    &110    &110\\ \hline
   \multicolumn{6}{|c|}{ provinces: $N_{c,p}$ }    \\ \hline
 $\kappa$   & 62.41&    61.07   &   61.07   &   61.07   &   61.08   \\
  $\chi$    & 0.369 &   0.371   &   0.371   &   0.371   &   0.371       \\
  $R^2$ & 0.973 & 0.985     & 0.985 & 0.985 & 0.985\\\hline
    \multicolumn{6}{|c|}{regions:  $N_{c,r}$   }    \\ \hline
     $\kappa$   & 225.97&   225.56  &   225.56& 225.77  &   225.77  \\
  $\chi$    & 0.607 &   0.608   &   0.608   &   0.608   &   0.608   \\
  $R^2$ & 0.953 & 0.953     & 0.953 & 0.953 & 0.953\\ \hline
\end{tabular}
\caption{Parameters of the Lavalette function, Eq.
(\ref{Lavalette2}), for the  fits  (see  data displayed in Fig.
\ref{fig:Plot4liloNcp3fNcr3f}) of the number of cities  in regions
and in provinces,  for  various years; the number of regions $N_r$
is always  equal to 20; the number of provinces  $N_p$ has  changed
as indicated.}\label{TableprovLavfitt}
\end{center} \end{table}

\begin{table} \begin{center}
\begin{tabular}[t]{cccc}
\hline &$N_{c,p}$&$N_{c,r}$\\
  \hline
Minimum &   6   &   74  \\
Maximum &   315 &   1544    \\
Mean     ($\mu$) &  73.564  &   404.6   \\
Median  ($m$)&  60  &   319 \\
RMS &   91.902      &   536.998     \\
Std Deviation    ($\sigma$)&    55.338  &   362.253     \\
Variance    &   3 062.27    &   131 227.52  \\
Std Error   &   5.2762  &   81.0023     \\
Skewness    &   1.7294 &    2.1284\\
Kurtosis    &   3.6845 &    3.8693\\
  \hline
 $\mu/\sigma$ &1.329   & 1.117 &\\
$3(\mu-m)/\sigma$ &0.7353&0.7089 &   \\
\hline
\end{tabular}
\caption{Summary of  (rounded) statistical characteristics for the
number ($N_c=8092$) distribution of IT cities in the various
($N_p=110$) provinces and regions   ($N_r=20$) in 2011. The maximum
$N_{c,p}$ value is 315 for (TO), while the minimum one is 6 for
(TS);  $N_{c,r}=$ 1544  (Lombardia) and 74  (Valle  d'Aosta)  for the
regions respectively, - see Table \ref{TableNcityperregion}.}
\label{Tablestatcityperregion}
\end{center} \end{table}

\begin{table} \begin{center}
\begin{tabular}[t]{ccc}
  \hline  &$N_{c,r}$ \\ \hline
 Lombardia& 1544\\
Piemonte    &1206\\
Veneto  &581\\
Campania&   551\\
Calabria&   409\\
Sicilia&    390\\
Lazio   &378\\
Sardegna    &377\\
Emilia Romagna&    348\\
Trentino Alto Adige&    333\\
Abruzzo&    305\\
Toscana &287\\
Puglia& 258\\
Marche  &239\\
Liguria&    235\\
Friuli Venezia Giulia&  218\\
Molise  &136\\
Basilicata  &131\\
Umbria& 92\\
Valle d'Aosta&  74\\  \hline
\end{tabular}
\caption{Number $N$ of (8092) cities (in 2011) in  the (20) IT
regions;  such a region ranking by city number corresponds to that
illustrated in Figs. \ref{fig:Plotnogood9ZMPpwldexp}-
\ref{fig:Plot4liloNcp3fNcr3f}. } \label{TableNcityperregion}
\end{center} \end{table}
The best Lavalette 2-parameter fits, with Eq. (\ref{Lavalette2_106})
form, are found in Table \ref{TableprovLavfitt}. Some illustrative
statistical characteristics of the city distributions as function of
region $r$ and province $p$,
 $N_{c,r}$ and $N_{c,p}$ respectively,
- in 2011 as an example,  are also given in Table \ref{Tablestatcityperregion}.


\begin{table}[htdp]
\begin{center}
\begin{tabular}{|c|c|c|c|c|c|}
\hline
  \multicolumn{3}{|c|}{ 2007}& \multicolumn{3}{|c|}{ 2008}  \\ \hline \hline
 Altidona &(AP)&  29 235 733& Altidona &(FM)&    30 329 015\\
\hline
 Andria &(BA)    &565 869 043&       Andria& (BT)    &581 635 172\\
\hline
 Arcore& (MI)&  293 056 037&    Arcore& (MB)&  300 146 626\\
\hline
 Arzana &(NU)&   17 002 253&     Arzana& (OG)&   18 200 141\\
 \hline
\end{tabular}
\caption{Examples of 4 cities, and their
 ATI, - observe quite different orders of magnitude, having a province
change but remaining in the same region, at their years change. Data
are expressed in Euros.}\label{7a}
\end{center}
\end{table}

\begin{table}[htdp]
\begin{center}
\begin{tabular}{|c|c|c|c|}
\hline
 \multicolumn{2}{|c|}{ 2007 (PU) - Marche}&  \multicolumn{2}{|c|}{ 2008 (RN) - Emilia Romagna}  \\ \hline
 Casteldelci   &       3 221 694&    Casteldelci & 3 171 730\\
   Maiolo     &         7 395 158 & Maiolo &       7 596 247\\
 Novafeltria  &        78 547 921   &Novafeltria &    80 178 021\\
 Pennabilli  &        28 814 429     &Pennabilli & 29 100 286\\
 San Leo  &         27 411 857  &San Leo &      28 792 554\\
 St Agata Feltria  & 24 563 898&     St Agata Feltria &24 046 727\\
 Talamello  &       11 371 705   &Talamello &       11 808 818\\
 \hline
\end{tabular}
\caption{The 7 cities (see text) having had a province change and
also a region change; their ATI is given at  their years change. As
written in the Table, PU (the province of Pesaro and Urbino) is in
the Marche region, while RN (province of Rimini) is in the Emilia
Romagna region. Data are expressed in Euros.} \label{7b}
\end{center} \end{table}

\subsection{Regional disparities}\label{regionaldisparities}

In this section, in view of respecting  "scientific constraints"
which impose to tie geography and economy along New Economy
Geography ideas (Krugman 1995), we consider every IT region  (made
of provinces and cities). We search whether the ATI of the cities in
each region  obey simple hierarchical relationships, - like a
2-parameter free Lavalette function. \newline First of all, it is
worth to point out that 228  municipalities have changed from a
province to another one, but nevertheless remained in the same
region (see Table \ref{7a} for a few examples), while 7
municipalities have changed from a province to another one, -in fact
also changing from a region to another (these 7 cases are given in
Table \ref{7b}). \newline Therefore, one can summarize the number of
cities belonging to a region as in Table \ref{TableNcityperregion}.
 This corresponds to
Figs. \ref{fig:Plot2Ncplilof}-\ref{fig:Plot4liloNcp3fNcr3f},
 in fact.
The display of  the distribution characteristics of these cities for
the 110 provinces  obviously requests  110  Tables (or Figures).
They are not given here, but any province case can be available from
the authors, -  upon request. \newline The following points have to
be taken into account before display and analysis:

\begin{itemize}
\item[(i)] the plot
illustrating the relationship between $N_{c,r}$ (and $N_{c,p}$)  and
their respective rank is year dependent;

\item[(ii)] the same comment applies for
ATI$_{c,r}$ (and ATI$_{c,p}$),  in obvious notations: they are year
dependent;
\item[(iii)]  finally, it is worth noting that the plots
of the relationship between the ATI$_c$, i.e. aggregated to the whole country,  and their rank is year
dependent, but not due to the change in the number of cities. This
simplifies the analysis.
\end{itemize}

A technical point is needed here. In order to optimize   the
fit procedure, i.e., also in order to have a  $\kappa$  value  characterized by a
few digits, the Lavalette function, Eq. (\ref{Lavalette2})   has
been   thereafter  opportunely rescaled by a $10^6$ factor  ($\sim y(N/2)$) also dropping the $N$ factor of  the rank $r$:
\begin{equation} \label{Lavalette2_106}
y(rank)= \hat{\kappa}\;10^6 \Big(\frac{rank}{N-rank+1}\Big)^{-\chi}.
\end{equation}

\subsection{ATI distributions in IT regions. Time, "King", and
"Vice-Roy Effects"}\label{ATIregions}

\begin{figure}
 \includegraphics[height=8.8cm,width=10.8cm]   {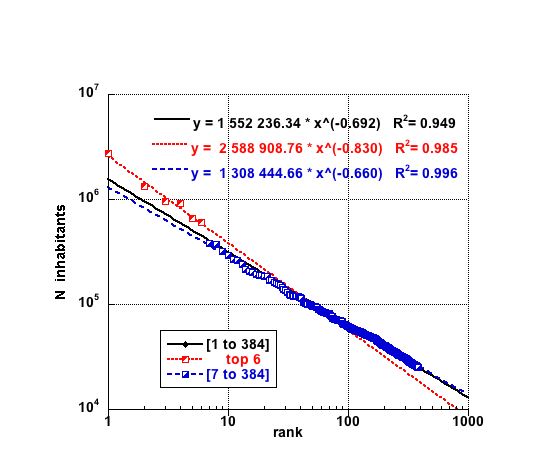}
\caption{The 384 largest IT cities ranked by decreasing order
according to their population size with corresponding power law fits
as indicated, pointing to 6 outliers.}
\label{fig6a:plotbigITcitieslolo}
\end{figure}

\begin{figure}
\includegraphics[height=8.8cm,width=10.8cm]{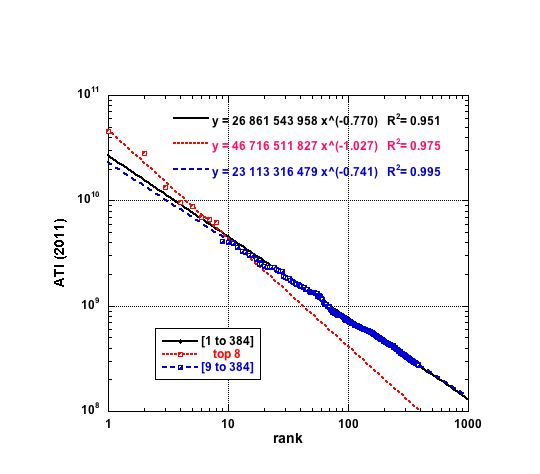}
\caption{  The 384 "richest"  IT cities ranked by decreasing order
according to their ATI with corresponding power law fits as
indicated, pointing to 8 outliers.} \label{Plot17ATI011top384}
\end{figure}

Before displaying and discussing the evolution of the various
regions from the ATI of their member cities point of view, a
practical remark is in order.  It is often found, and has been found
in the present study, that an upsurge occurs at low ranks. In other
words, the best (simplest, like power law or exponential or Zipf, as
those considered in Sect.\ref{citydistributions})  fits are impaired
because the low rank  data can be much above  (sometimes an order of
magnitude) whatever function is used in the appropriate fit,
resulting in an outlier for $r\rightarrow1$. This, observed a long
time ago by Jefferson (1989),  has been called a \textit{king
effect} by Laherrere and Sornette (1998), when examining the
population size of French cities (or rather agglomerations). For
example, the number of inhabitants in Paris is much bigger than the
(theoretical)  value resulting from the best (estimated, stretched
exponential) plot. In presence of only one outlier, the king  (K)
effect is identified. When an occurrence of several outliers is
observed, then there is \textit{king plus vice-roy effect} (KVR).
\newline
Such ATI  (or city) outliers are observed  in almost all regions and
provinces, as shown below.
\newline
For convincing the reader, let two cases be shown, as examples:
\begin{itemize}
\item consider  the 384 largest
IT cities, in terms of population size\footnote{Population refers to
the Census 2011 data.},  for the whole Italy, as ranked by
decreasing order, and compare such a size-rank relationship   to a
power law;  as indicated in Fig. \ref{fig6a:plotbigITcitieslolo}, it
is obvious that there are 6 "outliers" (in order from the biggest:
Roma, Milano, Napoli, Torino, Palermo, Genova);
\item a similar situation occurs when
examining ATI values, rather than population sizes:  consider  the
384 "richest" IT cities, in terms of ATI size,  for the whole Italy,
as ranked by decreasing order, and compare such a size-rank
relationship   to a power law;   see   Fig.
\ref{Plot17ATI011top384}, it is obvious that there are 8 "outliers"
(in order from the biggest: Roma (RM), Milano (MI), Torino (TO),
Genova (GE), Napoli (NA), Bologna (BO), Palermo (PA), Firenze (FI)).
For completeness,    let it be known that from the ATI ranking point
of view, the top 12 IT cities have never changed their ranking, i.e.
these 8 plus Venezia (VE), Verona (VR),  Bari (BA), and Padova (PD).

 \end{itemize}

Observe that $6\neq8$,  see  that cities are differently ranked, and
what city is added to the ATI outliers with respect to the
population size ones.
\newline
Although the demonstration in such figures is made through a log-log
plot with power law fits, the same effects occur when using
exponential or Lavalette function fits on semi-log plots. Similar
situations occur for the regional and provincial level though not
necessarily so well marked due to the smaller number of data points
and their size value, - surely in the province cases. Nevertheless,
in order to obtain some reasonable estimates of the empirical
relations over a large range of data, it seems obviously necessary
to take into account such a king effect, - in almost all the data,
we have examined. Moreover, because such king effects, as seen in
Figs. \ref{fig6a:plotbigITcitieslolo} - \ref{Plot17ATI011top384}, in
fact truly occur over a rank interval $\ge 1$, it has been necessary
to consider king plus vice-roy effect, accounting for more than 1
outlier, - as made more precise in the figure captions.
\newline
When a flattening of the data occurs at low rank, a so called
\textit{queen}, or often a  \textit{queen plus harem}, effect
appears (Ausloos 2013); the  "problem" is  different  from the KVR
effect; a Zipf-Mandelbrot-Pareto law is of course a more appropriate
description, in such cases.  None has   been found  to occur  in the
present study.
\newline
Nevertheless, a special case has to be pointed out at once here.
Although, it is shown that the Lavalette law usually well represent
the ATI data,   a 3-parameter Zipf-Mandlebrot-Pareto law fits
unexpectedly well  the Lazio region data, - as long as the rank is
$2\le r \le 101$. The illustration, statistical analysis and some
specific discussion  are postponed  to Appendix A, for this special
region.  This  finding confirms a classical statement, i.e. the
soundness of the Zipf's law can hold  for a subset  of a collection
of data, but  does not necessarily hold for the entire set. This is
in accord with the empirical evidence registered in previous studies
(see Section \ref{Review},  for  references to the literature on
this).
\newline
Results are displayed in Figs.
\ref{newgreenplots/Plot6Abruzzo}-\ref{fig:Plot6Venetolilo5f2Lg},
whose captions are rather detailed. The parameters of the best fits
are reported in Tables
\ref{TableIATIregionfits}-\ref{TableIIATIregionfits}. A discussion
is presented in Section \ref{sec:results}.

\section{Results and discussion}\label{sec:results}

This section fixes and discusses the results of the investigation.
\newline
 First, a rank-size rule, on the basis of the
number of cities per province, has been  searched through
Zipf-Mandelbrot-Pareto, power and exponential laws. It statistically
failed. However, the rank-size rule for the cities in Lazio region
can be well described by those curves. This fact confirms the
finding of some researchers that a subset of a sample can be well
represented by Zipf's law while the whole sample may fail in this
doing (we address the reader to the discussion in Section
\ref{Review} and Appendix A). Should it be necessary to the reader
to recall that the Lazio region contains Roma, the capital city of
Italy? and  can thus  be expected to present a superking effect.
\newline The 2-parameter Lavalette law seems to suitably fit, -
with a high level of $R^2$ and/or visual soundness between curve and
data, the rank-size rule for Italy cities under different
perspective and size-detection criteria. Specifically: $(i)$ number
of cities per region; $(ii)$ number of cities per  province. The
occurring deviations for low rank, more evident in case $(ii)$, are
due to the  (KVR-like) outliers and to the creation of 7 new
provinces during the observed period.  \newline In exploring the
regional cases, several facts emerge. \newline As for what concerns
the low-rank elements in the Zipf's law case (see e.g. Gabaix,
2009), the role of the outliers at high rank is rather huge in the
Lavalette case. For several regions, a strong king or king plus
vice-roy effect may destroy the statistical consistence of the mere
2-parameter Lavalette curve in plotting the data. The $R^2$ is not
necessarily small, the visual appeal of the fit is weak: this is due
in such fits to the importance taken by the low rank (thus high ATI
values) of  a few cities. In such cases, removing the outliers can
lead to a more convincing fit (paradigmatic cases are Aosta Valley,
Basilicata, Campania, Friuli Venezia Giulia, Liguria, Lombardia,
Molise, Puglia, Sicilia, and Trentino Alto Adige). Other cases
provide a substantial indifference in removing the outliers, with
neither an appreciable improvement of the visual appeal of the
graphs nor of the $R^2$ (many cases are not displayed, for
shortening the  paper),
 like Abruzzo, Marche, Sardegna, Umbria and Veneto.
A  few cases give rise to   questions, but with some  answer: in
fact, in several  cases the removal of the outliers implies
unexpected  not much better results from a
 $R^2$, point of view, but  in presence of a better visualization of the fit; this is the case of
Friuli Venezia Giulia. A slightly less appealing visualization of
the fit with a slightly smaller $R^2$ occurs also for Emilia
Romagna. Valle d'Aosta is the region where the KVR-effect must be
removed for a fine fit.
\newline
Sometimes, there is some surprise, thus no real "answer": Trentino
Alto Adige, Molise and Sicilia are found to have a large number of
vice-roys. Also, fits to the Marche data are rather insensitive to a
KVR effect removal, although the $R^2$ is {\it at first}, for the
raw data, not very high.
\newline
Finally, Lazio seems to be not properly described by a 2-parameter
Lavalette function, but rather through exponential, ZMP, and power
laws, as already mentioned (see the discussion above).
\newline
In view of the above, it seems that there is some evidence that the
KVR effects are not due to scale factors, but are  intrinsic to the
regularities and discrepancies, since the KVR effect occurs in most
cases, -  found in quite different size systems.

\section{Conclusions}\label{conclusions}
This paper provides a statistical analysis of the Italian
municipalities for the period 2007-2011, ranked by their ATI values.
It is proven that while ZMP, exponential and power laws are
not statistically appealing in describing the size-rank rule, a
2-parameter Lavalette function is. To the best of our knowledge,
this is the first time that such typology of function is employed in
urban studies.
\newline
Data also confirm that IT is a unique entity, but with different
regional realities. Several  cities play a prominent role in
determining the Italian GDP;  they are detected within the regions
through the king and king plus vice-roy effects. We have observed
that there is some evidence that the KVR effects are not due to
scale factors, but are  intrinsic to the  economic regularities and
discrepancies.
\newline
A few cases are puzzling, and suggest some further investigation of
this theme. Thus, a refinement of the analysis through the
introduction of a 3-parameter Lavalette function or a modified
version of it is in order. In particular, the second aspect suggests
to work in the direction of a theoretical improvement of the current
literature on the laws describing rank-size rules.


 \clearpage

 \newpage

 \section*{Appendix A. The Lazio case}

\begin{figure}
\includegraphics[height=8.9cm,width=9.8cm] 
{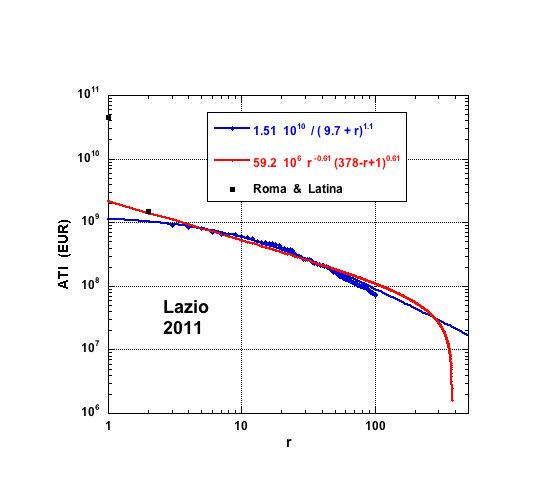}
 \caption   {Log-log plot  of the  ranked 2011 ATI Lazio cities -$r$ represents here the rank-,
showing fits by a Lavalette function (red line)  and a
Zipf-Mandelbrot-Pareto function (blue line), for  $3\le r \le 101$,
i.e.  when the  king (Roma) and vice-roy (Latina) data points (large
black square dots) are excluded from the fits.}
\label{Plot37ZMPsogoodLavLazio100}
\end{figure}

\begin{figure}
\includegraphics[height=7.4cm,width=9.8cm] 
{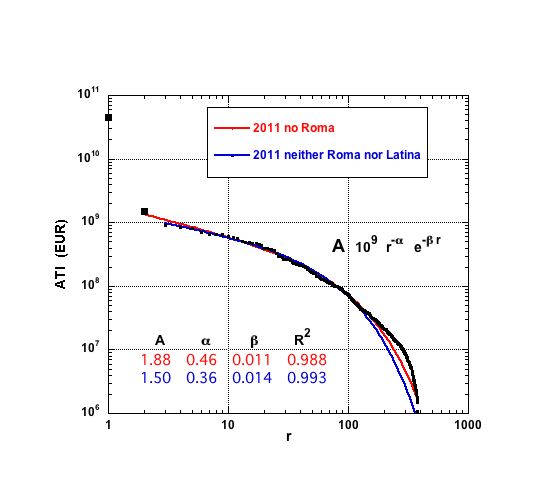}
 \caption   {Log-log plot  of the  ($r$) ranked 2011 ATI Lazio cities, showing fits by a power law with an exponential  cut-off, when either (red line) the king (Roma) or  (blue line) king plus vice-roy (Latina) data points  (large black square dots) are excluded from the fits; the regression
coefficients are given.} \label{Plot28Lazio11noRnoLpwlco}
\end{figure}

\begin{figure}
\includegraphics[height=7.4cm,width=9.8cm] 
{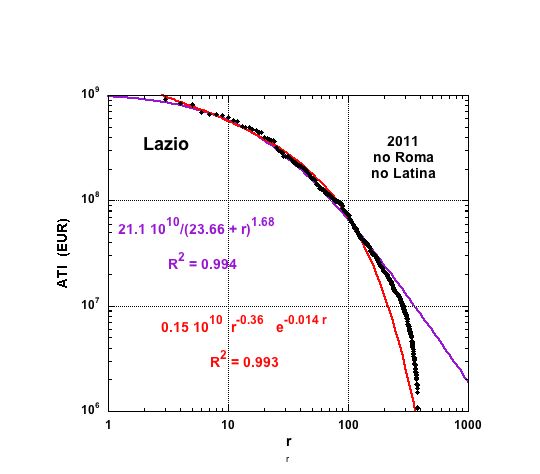}
 \caption   {Log-log plot
of the  ranked 2011 ATI Lazio cities, showing fits by a power law
with an exponential  cut-off  (red line) at high rank ($r$) compared
to a ZMP law (violet line), with their corresponding regression
coefficient, when either the king (Roma) and vice-roy (Latina) data
points  are excluded from the fits.} \label{Plot11Lazio11mixnoKVR}
\end{figure}

It has been indicated in the main text that a 3-parameter
Zipf-Mandelbrot-Pareto law fits unexpectedly well  the Lazio region
ATI data, - as long as the rank is $3\le r \le 101$, much better
than a Lavalette function; see Fig. \ref{Plot37ZMPsogoodLavLazio100}
for the 2011 case, on a log-log plot (and Fig. \ref{fig:Plot45LazionoKVRspecialmix} for all 5 years on a semi-log plot). Another mere exponential fit   (not shown) to the
whole data, except for the king (Roma) and vice-roy (Latina) data
points also  indicates a strong cut-off at $r\ge100$.
\newline
For illustration and completeness, indicating that other possible
fits were investigated, Figs.
\ref{Plot28Lazio11noRnoLpwlco}-\ref{Plot11Lazio11mixnoKVR} show   a
fit to a power law with exponential cut-off at high rank,  and a
comparison of such a fit with a Zipf-Mandelbrot-Pareto law,
respectively, on a log-log plot, for the ATI 2011 year, - as an
example, when either  the Roma point or the Roma and Latina data
points are not considered.

 \clearpage

 \section*{Appendix B. Tables and Figures}

\begin{figure}
\includegraphics[height=8.0cm,width=9.8cm]{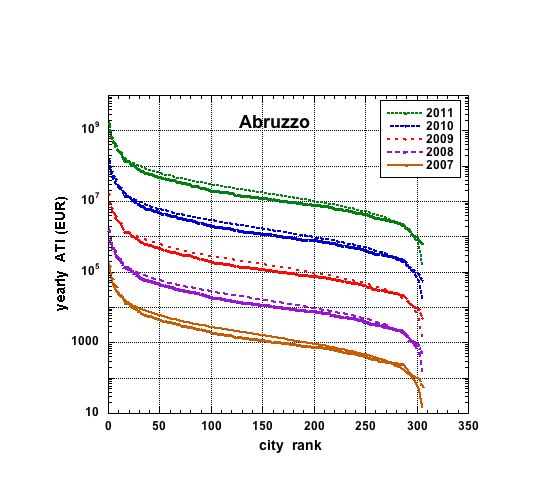}
\caption   {Abruzzo, a regional case of ATI distributions; $N$=305
cities are ranked accordingly;  with 2-parameter Lavalette fits;
neither king plus vice roy effect nor king effect is observed.} 
  \label{newgreenplots/Plot6Abruzzo}
\end{figure}

\begin{figure}
\includegraphics[height=6.0cm,width=9.8cm] {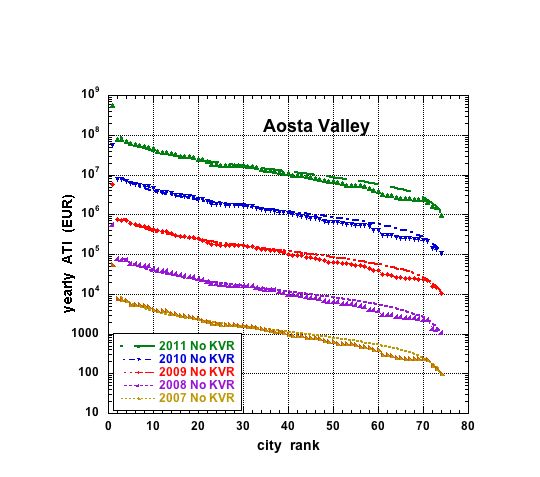}
\caption   {Ranked ATI yearly values  for Aosta Valley, a  regional
case with (obviously)   king plus vice-roy effect (Aosta and Sarre);
$N$=74; with 2-parameter Lavalette fits. 
} \label{fig:Plot12Aostalilo5f2Lbad}
\end{figure}

\begin{figure}
\includegraphics[height=6.0cm,width=9.8cm] {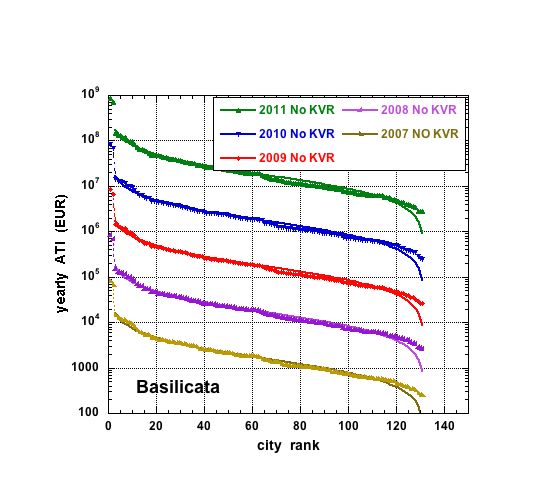}
\caption   { Fit of a 2-parameter free Lavalette function to ranked
ATI yearly values  for the $N=131$ ranked cities, when removing
Potenza and Matera, as king and vice-roy cities.}
\label{fig:Plot12Basilicatano2kings}
\end{figure}

\begin{figure}
\includegraphics[height=6.0cm,width=9.8cm] {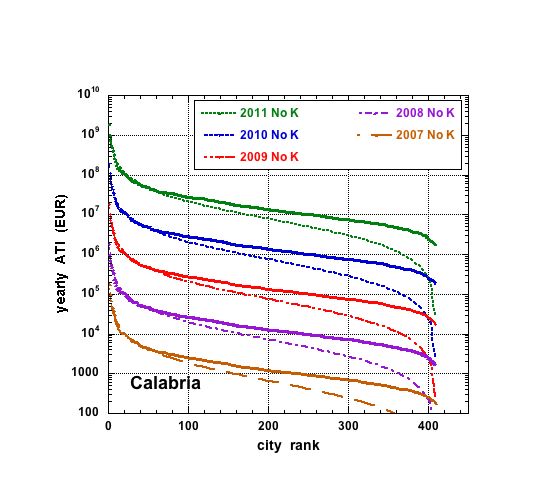}
\caption   {Calabria city yearly ATI ranking distribution
2-parameter Lavalette fits; $N$=409, but removing   a remarkable
king effect as Reggio Calabria   for the fit. Nevertheless, note the
departure from a "good looking fit" at high rank in the most recent
years.} \label{fig:Plot30Calabria5f2L}
\end{figure}

\begin{figure}
\includegraphics[height=8.0cm,width=9.8cm] {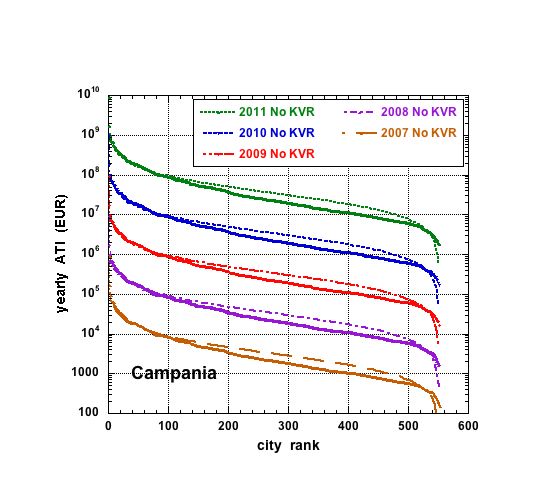}
\caption   {Campania,  city yearly ATI ranking distribution and
2-parameter Lavalette fits ($N$=451), but  after removing   a
remarkable king plus vice-roy effect (Napoli and Salerno).}
\label{fig:Plot2Campania2f2L}
\end{figure}

\begin{figure}
\includegraphics[height=8.0cm,width=9.8cm] {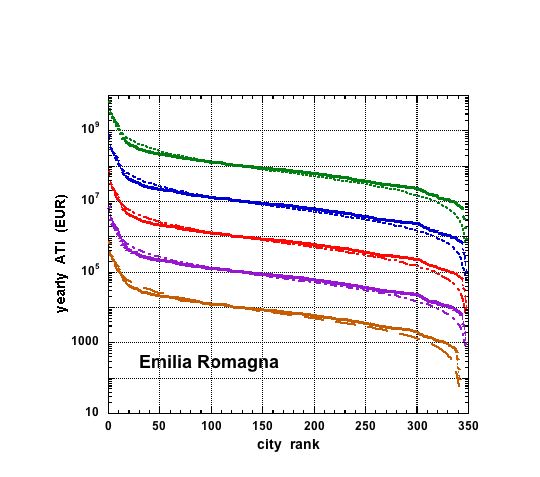}
\caption   { Ranked ATI yearly values  for Emilia Romagna cities
with 2-parameter Lavalette fits. N.B. $N=341$ in 2007, but $N=348$
otherwise. Moreover, there is no need for optimizing the fit in
considering Bologna  as inducing a king effect. Nevertheless, note
the departure from a "good looking fit" at high rank in the most
recent years.} \label{fig:Plot10EmilRom4f2L}
\end{figure}

\begin{figure}
\includegraphics[height=8.0cm,width=9.8cm] {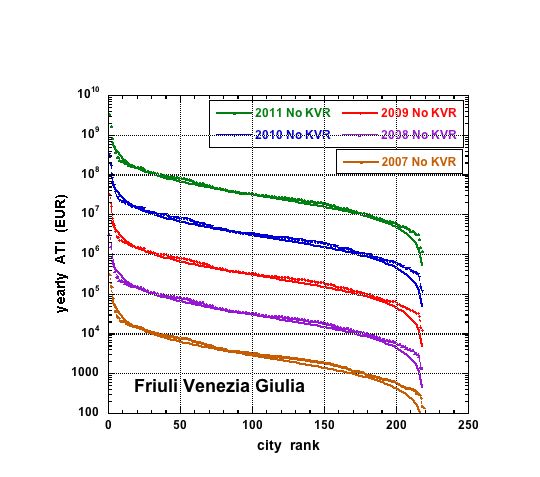}
\caption   {Friuli Venezia Giulia city ATI distribution:   $N$=219
in 2007 $\rightarrow$ 218 thereafter and 2-parameter Lavalette fits,
but admitting a king and vice-roy  (Trieste and Udine) effect, - as
observed when such a  fit on  the full data is attempted.}
\label{fig:Plot5FriuliVGlilo0113f2L}
\end{figure}

\begin{figure}
\includegraphics[height=8.0cm,width=9.8cm] {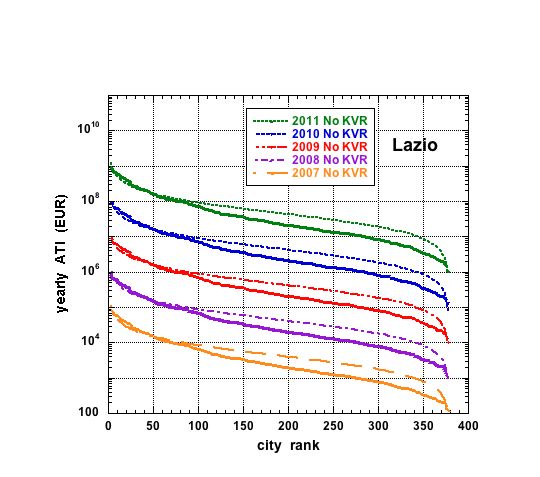}
\caption   {ATI yearly distribution for Lazio   ($N$=378) cities;
2-parameter Lavalette fits, after removal of   king plus vice-roy
effect (Rome and Latina).  }   \label{fig:Plot49LaziolilonoRnoL2f2L}
\end{figure}

\begin{figure}
\includegraphics[height=8.0cm,width=9.8cm] {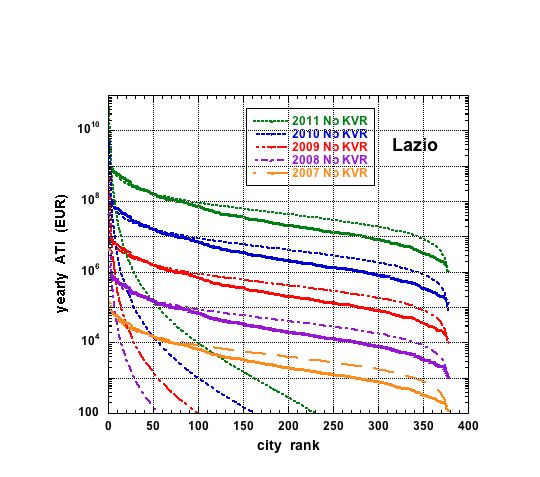}
\caption   {Lazio,  $N$=378 cities; 2-parameter Lavalette fits.
Comparison between raw data and removal of king plus vice-roy effect
(Rome and Latina) is amazing, - in this worse encountered case.}
\label{fig:Plot45LazionoKVRspecialmix}
\end{figure}

\begin{figure}
\includegraphics[height=8.0cm,width=9.8cm] {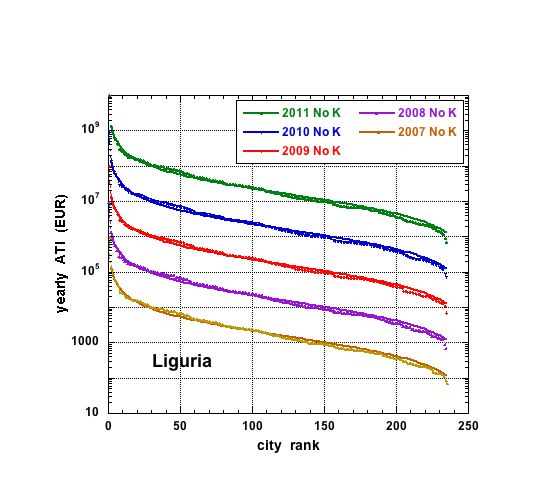}
\caption   { Ranked ATI yearly values  for   Liguria ($N$=235)
cities; 2-parameter Lavalette fits, after removal of   king  effect
(Genova).}   \label{fig:Plot1Ligurialilo2f2LK}
\end{figure}

\begin{figure}
\includegraphics[height=8.0cm,width=9.8cm] {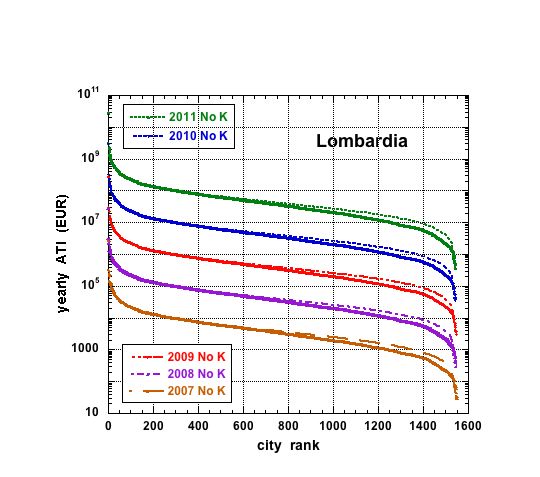}
\caption   { Ranked city ATI yearly values  for  Lombardia region:
2010 and 2011 for $N$=1546 cities;  2007, 2008, and 2009  for
$N$=1544 cities; fits with the 2-parameter Lavalette function, after
removing the Milano king effect data point.}
 \label{fig:Plot74Lombardia3f2LKg}
\end{figure}

\begin{figure}
\includegraphics[height=8.0cm,width=9.8cm] {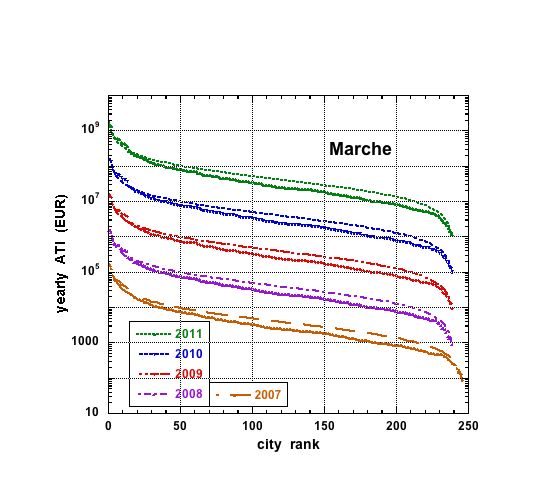}
\caption   {Marche:  $N$=246 cities in 2007 and $N$=239 thereafter;
fits of ATI  yearly  raw data with the 2-parameter Lavalette
function. No king or king plus vice-roy effect is observed  in this
region.}
 \label{fig:Plot68Marchelilo5f2L}
\end{figure}

\begin{figure}
\includegraphics[height=8.0cm,width=9.8cm] {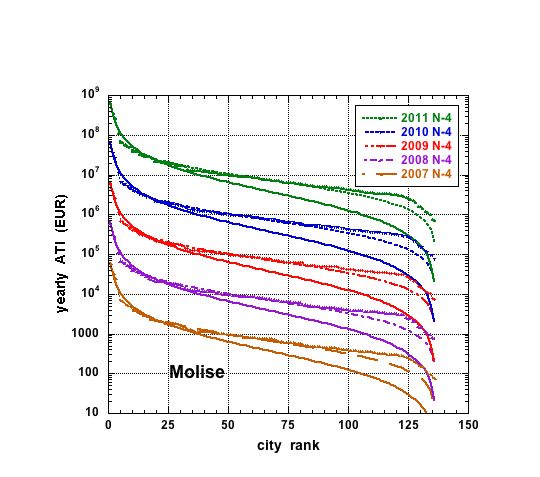}
\caption   {Molise $N$=136 cities ranked according to their yearly
ATI, with the 2-parameter Lavalette function fits. Remarkably a king
(Campobasso) with 3 vice-roys (Termoli, Isernia, Venafro), i.e. N-4
data points are used, effect here is very  meaningful; the
corresponding fits on the whole  $N$=136  data are indicated by
continuous lines.} \label{fig:Plot7Moliselilo5f2L}
\end{figure}

\begin{figure}
\includegraphics[height=8.0cm,width=9.8cm] {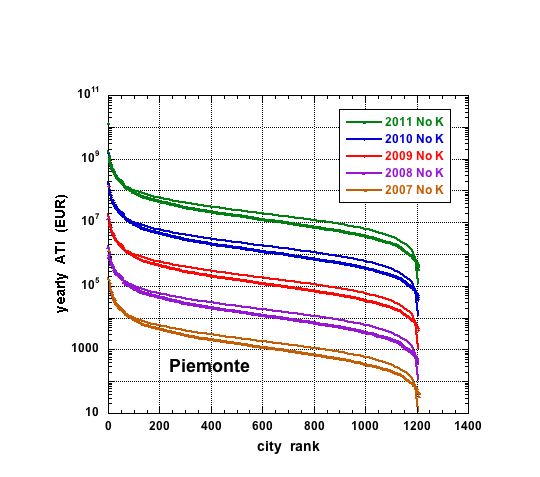}
\caption   {Piemonte   $N$=1206 cities yearly ATI with 2-parameter
Lavalette function fits. Remarkably a king (Torino) must be
withdrawn for a  realistic  fit  improvement }
\label{fig:Plot2Piemonte2f2LKg}
\end{figure}

\begin{figure}
\includegraphics[height=8.0cm,width=9.8cm] {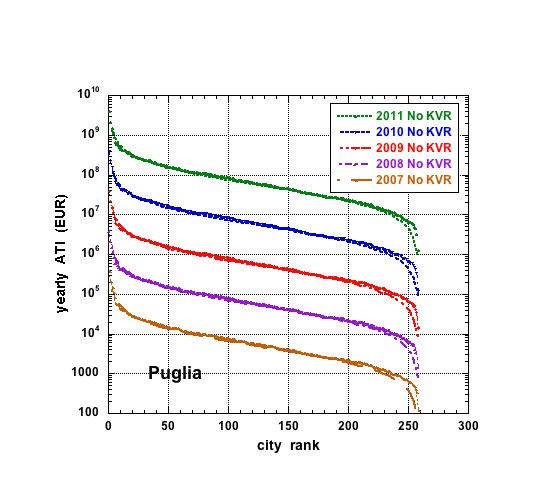}
\caption   {Yearly ATI  ranked data of Puglia  $N$=258 cities
fitted with the 2-parameter Lavalette function,  taking into account
a  king  (Bari) and a vice-roy (Taranto) effect, - for  very fine
fits} \label{fig:Plot4Puglialilo5f2L}
\end{figure}

\begin{figure}
\includegraphics[height=8.0cm,width=9.8cm] {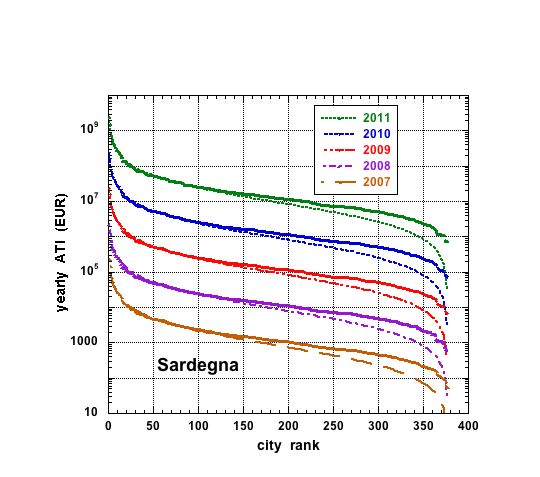}
\caption   {Yearly ATI ranked data of Sardegna,  $N$=377 cities,
with 2-parameter Lavalette function for fitting.  Considering any
king or king plus vice-roy effect gives not  much improvement of the
fit.} \label{fig:Plot6Sardegna5f2L}
\end{figure}

\begin{figure}
\includegraphics[height=8.0cm,width=9.8cm] {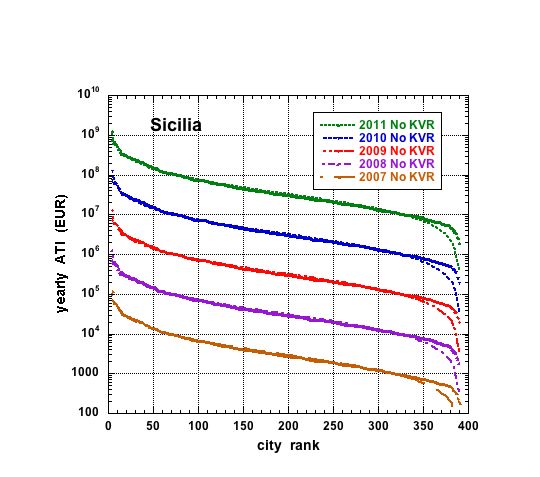}
\caption{ Ranked ATI yearly value distributions for Sicilia cities:
$N$=390 and subsequent 2-parameter Lavalette fits, but admitting a
king and two  vice-roys  (Palermo, Catania and Messina) effect.}
\label{fig:Plot72Sicilialilo5f2Lg}
\end{figure}

\begin{figure}
\includegraphics[height=8.0cm,width=9.8cm] {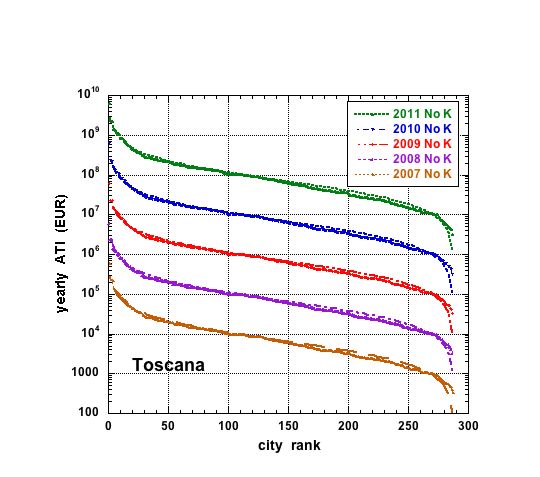}
\caption   {Toscana city yearly ranked ATI distribution:   $N$=287
and subsequent
  2-parameter Lavalette fits, but admitting a king   (Firenze) effect.}
\label{fig:Plot3Toscanalilo5f2Lg}
\end{figure}

\begin{figure}
\includegraphics[height=8.0cm,width=9.8cm] {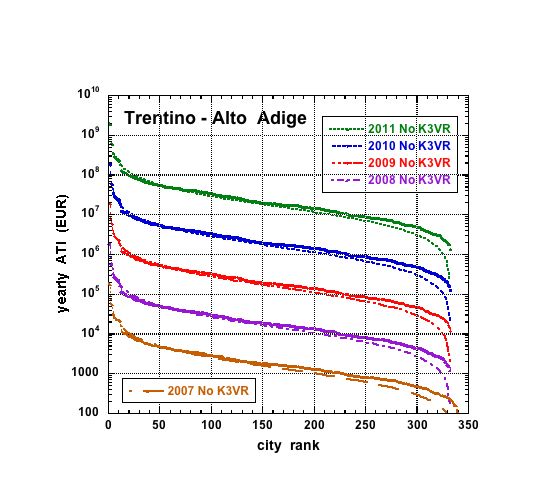}
\caption{ Trentino-Alto Adige: a regional case comparing ATI values
through 2-parameter Lavalette fits in 2008-2011 for $N$=333 cities,
and in 2007  for $N$=339 cities, taking into account in both cases a
king (Trento) and three vice-roys (Bolzano, Merano, and Rovereto),
thus removing  the corresponding data points before fits.}
\label{fig:Plot20TAAdigelilo4f2L}
\end{figure}

\begin{figure}
\includegraphics[height=8.0cm,width=9.8cm]{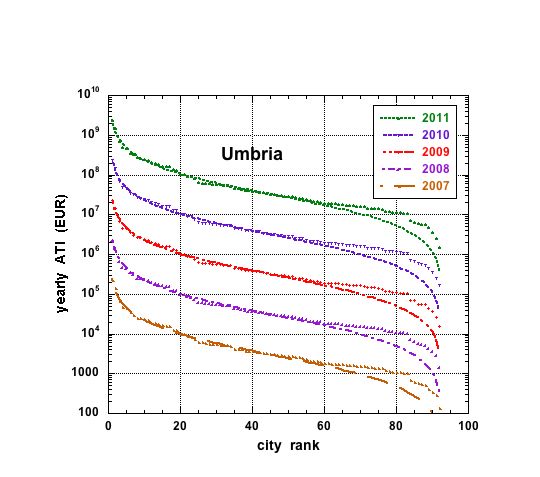}
\caption{Distribution of Umbria  $N$=287 cities ranked according to
their yearly ATI,  - data fitted with the 2-parameter Lavalette
function. No need to search for king or king plus vice-roy effect,
but  note the remarkable hump at $r$=60 with departure from the fit
at high $r$.} \label{fig:Plot5Umbrialilo5f2Lg}
\end{figure}

\begin{figure}
\includegraphics[height=8.0cm,width=9.8cm]{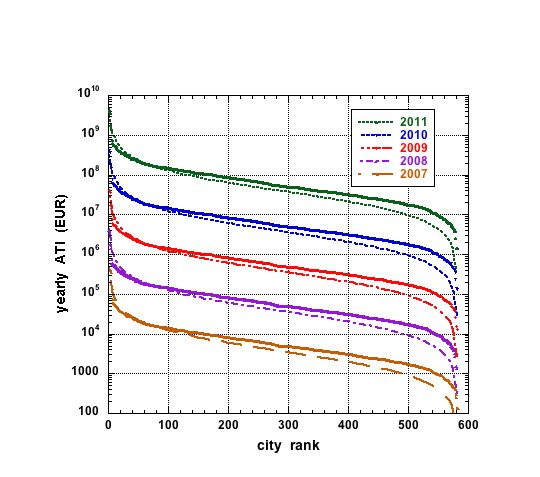}
\caption{Distribution of Veneto  $N$=581 cities ranked according to
their yearly ATI,  - data fitted with the 2-parameter Lavalette
function. No need to search for king or king plus vice-roy effect. }
\label{fig:Plot6Venetolilo5f2Lg}
\end{figure}

This Appendix contains
\begin{itemize}
\item[(i)]  the figures,  Figs.
\ref{newgreenplots/Plot6Abruzzo}-\ref{fig:Plot6Venetolilo5f2Lg},
relative to the  ranking of  cities according to their ATI, and best
fits by a Lavalette function, sometimes for the raw data, sometimes
taking into account a K  effect of a   KVR effect.   It is here
mentioned, once and for all, that the data 2011 data is $not$
rescaled, but all ATI data scales for the other years are
systematically reduced for the display  by a factor 10$^m$, where
$m$ is the difference between 2011 and the year of interest;
\item[(ii)] the parameters of the best fits to a Lavalette function of the raw data,  in Tables
\ref{TableIATIregionfits}-\ref{TableIIATIregionfits};  a column
indicates how many KVR cities can be considered (and removed) in
order to optimize the fits reported in the corresponding figures;

\end{itemize}

\clearpage

\clearpage

 \begin{table}   \begin{center}
 \begin{tabular}{|c|c|c|c|c|c|c|c|c|}
   \hline
 Region  & &  2007 & 2008&2009&2010& 2011& KVR &Fig. \\\hline
Abruzzo&$\hat{\kappa}$&15.43        &15.725     &15.89      &16.19      &16.81 & &  \ref{newgreenplots/Plot6Abruzzo}\\
Abruzzo&$\chi$&0.814    &0.805      &0.809      &   0.809   &0.805      & &\\
Abruzzo&$R^2$&0.986    &0.981      &0.986      &0.986 &0.986&0 &\\
\hline
Aosta Valley&$\hat{\kappa}$&    0.589   &   0.624   &0.635      &   0.665   &   0.698   &  & \ref{fig:Plot12Aostalilo5f2Lbad}\\
Aosta Valley&$\chi$&    1.574   &1.566  &1.566      &   1.558   &   1.546   &  &\\
Aosta Valley&$R^2$&    0.911&0.909     &   0.909   &   0.910   & 0.908&2 &
\\ \hline
Basilicata&$\hat{\kappa}$&7.223&7.782&7.888&7.866&7.782 & &\ref{fig:Plot12Basilicatano2kings}\\ Basilicata&$\chi$&0.978&0.966&  0.966&0.969&0.966   & &  \\
Basilicata&$R^2$&0.923&0.920&0.920&0.917&0.920  &2   &\\ \hline
Calabria&     $\hat{\kappa}$  &7.195&7.471& 7.834&7.947&8.103& & \ref{fig:Plot30Calabria5f2L}\\
Calabria&$\chi$&0.915&0.913  &0.909&0.907&0.902& &\\
Calabria& $R^2$&0.993&0.994& 0.994&0.994&0.994&1 &\\ \hline
Campania&$\hat{\kappa}$&0.134&0.151&0.166&0.169&0.151& &\ref {fig:Plot2Campania2f2L}\\
Campania&$\chi$&1.756&1.738&1.724&1.722&1.738& &\\
Campania&$R^2$&0.945&0.943&0.942&0.942&0.943&2 &\\  \hline
Em.Romagna(*) &$\hat{\kappa}$& 60.77&61.60&61.32&62.49&63.81& &\ref {fig:Plot10EmilRom4f2L} \\
Em.Romagna(*) &$\chi$& 0.810&0.807&0.807&0.804& 0.800 & &  \\
Em.Romagna(*) &$R^2$&
0.977&0.977&0.977&0.976&0.976&1 &\\
\hline
FriuliVG(**) &$\hat{\kappa}$&8.662&8.61&0.8219& 8.313&8.547& & \ref{fig:Plot5FriuliVGlilo0113f2L}\\
FriuliVG(**) &$\chi$&1.093& 1.099&1.110&1.108&   1.102& & \\
FriuliVG(**) &$R^2$&0.980&0.979&0.979&0.979&0.978&2 &\\
\hline
Lazio&   $\hat{\kappa}$    &       !&       !&      ! &  !    &    !   &   &\ref{fig:Plot49LaziolilonoRnoL2f2L}\\
Lazio&$\chi$&      !&      ! &       !&    !   &    !   &  &\\
Lazio& $R^2$&    !&  !     & !      & !      &    !   &2 &(\ref{fig:Plot45LazionoKVRspecialmix})\\\hline
 Liguria&   $\hat{\kappa}$         &0.028&0.030&0.030&0.031& 0.033& &\ref{fig:Plot1Ligurialilo2f2LK}\\
Liguria&$\chi$&2.327&2.321&2.321&2.317&2.307& &\\
Liguria& $R^2$&0.984& 0.984&0.984&0.984&0.983&1 &\\ \hline
Lombardia(***) &$\hat{\kappa}$&0.002&0.002 &0.002&0.002&0.002&  &\ref{fig:Plot74Lombardia3f2LKg}\\
Lombardia(***) &$\chi$&2.233&2.231&2.228 &2.243&2.253& &\\
Lombardia(***) &$R^2$&0.956&0.955&0.954&0.954&0.954&1 &\\\hline
\end{tabular}
\caption{ (I) Parameter values of the ATI data fits with the adapted 2-parameter  Lavalette function,
 Eq. (\ref{Lavalette2_106}),  every considered year:
$\hat{\kappa} *10^6*[r/(N-r+1)]^{- \chi }$;  with $\hat{\kappa}$= 1  and  $\chi$ =  1   as
initial iteration conditions; $N$ depends on the year and the
region:  it is usually given by the value in Table
\ref{TableNcityperregion}, except for (*) $N=341$ (2007)
$\rightarrow$ 348 (2008-11); (**) $N=219$ (2007) $\rightarrow$ 218
(2008-11); (***) $N=1546$ (2007-09) $\rightarrow$ 1544 (2010-11).
Lazio is so meaningless, see Fig. \ref{fig:Plot49LaziolilonoRnoL2f2L}, that values are not shown (see the
discussion in the text). KVR column stands for  how many
 king effect and king plus vice-roys effect     are taken into account in the mentioned  figure to improve  $R^2$$\rightarrow 0.99$.}
\label{TableIATIregionfits}
 \end{center}
 \end{table}

 \begin{table} \begin{center}
 \begin{tabular}{|c|c|c|c|c|c|c|c|c| }
   \hline
Region   & &  2007 & 2008&2009&2010& 2011& KVR&Fig
\\\hline
Marche($^{\#}$)  &$\hat{\kappa}$&     37.05&38.63&38.22&38.99&40.29& &\ref{fig:Plot68Marchelilo5f2L}\\
Marche($^{\#}$)  &$\chi$&0.696&0.695&0.697&0.695&0.689& &\\
Marche($^{\#}$)  &$R^2$&0.964&0.963&0.965&0.964&0.962& 0&\\\hline
Molise& $\hat{\kappa}$&3.525&3.672&3.539&3.552&3.605& &\ref{fig:Plot7Moliselilo5f2L}\\
Molise&$\chi$&1.049& 1.046&1.054&1.053&1.053&  &\\
Molise&$R^2$&0.979&0.978&0.979&0.978&0.978&3&\\\hline
Piemonte&  $\hat{\kappa}$  &0.005&0.005&0.006&0.006&0.007& &\ref{fig:Plot2Piemonte2f2LKg}\\
Piemonte&$\chi$& 2.092 &2.084& 2.062  &2.065&2.047& &\\
Piemonte& $R^2$& 0.952 &0.952 &0.951&0.950&0.949&1&\\  \hline
Puglia& $\hat{\kappa}$&34.33&  36.34&37.30 &37.87&39.29& &\ref{fig:Plot4Puglialilo5f2L}\\
Puglia&$\chi$&0.844&0.837&0.833&0.832&0.824 & &\\
Puglia&$R^2$&0.985&0.985 &0.985&0.985&0.985&2&\\\hline
Sardegna&  $\hat{\kappa}$ &8.141& 8.741&9.048&9.032&9.155&  &\ref{fig:Plot6Sardegna5f2L}\\
Sardegna&$\chi$&0.953&0.945&0.940&0.942&0.939& &\\
Sardegna&$R^2$& 0.986&0.987&0.987&0.987&0.988&0&\\\hline
Sicilia& $\hat{\kappa}$& 10.26&10.73&11.20& 11.18&11.71& & \ref{fig:Plot72Sicilialilo5f2Lg}\\
Sicilia&$\chi$&1.077&1.072 &1.067     &1.068    &1.058& & \\
Sicilia&$R^2$&0.983&0.982&0.982  &0.982   &0.982&3&\\   \hline
Toscana& $\hat{\kappa}$&47.39&48.47& 49.33&49.78&50.16& &\ref{fig:Plot3Toscanalilo5f2Lg} \\
Toscana&$\chi$&0.844&0.842&0.839& 0.839&0.839& &\\
Toscana&$R^2$&0.981&0.981&0.980&0.981&0.980&1&\\\hline
Tr.-A.Adige($^{\#\#}$)   &$\hat{\kappa}$&8.681&9.304&   9.573&9.982&9.304& & \ref{fig:Plot20TAAdigelilo4f2L}\\
Tr.A.Adige($^{\#\#}$)  &$\chi$&0.936&0.930&0.929&   0.924&0.930& &\\
Tr.-A.Adige($^{\#\#}$)  &$R^2$&
0.922&0.923&0.924&0.924&0.923&4&\\
\hline
Umbria& $\hat{\kappa}$&27.99&28.92&29.44&29.59&30.33& &\ref{fig:Plot5Umbrialilo5f2Lg}\\
Umbria&$\chi$&0.975&0.973&0.970&0.971&0.964& &\\
Umbria&$R^2$&0.986& 0.986 &0.986& 0.987&0.987&0&\\\hline
Veneto&$\hat{\kappa}$ &35.88&36.79&36.50&37.35&38.45&  &\ref{fig:Plot6Venetolilo5f2Lg}\\
Veneto&$\chi$&0.770&0.767&0.768&0.765&0.760& &\\
Veneto&$R^2$&0.895&0.895&0.896&0.895&0.897&0&\\   \hline
\end{tabular}
\caption{(II) Parameter values of the ATI data fits with the adapted 2-parameter  Lavalette function,
 Eq. (\ref{Lavalette2_106}),  every considered year:
$\hat{\kappa} *10^6*[r/(N-r+1)]^{- \chi }$;  with $\hat{\kappa}$= 1  and  $\chi$ =  1   as
initial iteration conditions; $N$ depends on the year and the
region:  it is usually given by the value in Table
\ref{TableNcityperregion}, except for : ($^\#$)  N= 246 in 2007
$\rightarrow$ 239 thereafter; ($^{\#\#}$) $N=336$ (2007) $\rightarrow$ 333
(2008-11).  KVR column stands for  how many
 king   and king plus vice-roys     are taken into account in the mentioned  figure to improve  $R^2$$\rightarrow 0.99$.}
\label{TableIIATIregionfits}
 \end{center}
 \end{table}

  \clearpage

\begin{landscape}
 \label{topandbottomcitiesATIranked}
    \begin{tabular}{clllllll}
       \hline
  \multicolumn{6}{|c|}{Top and bottom modifications of ranked IT cities according to their ATI,  during the 5 years of interest.}\\
   \multicolumn{6}{|c|}{Recall that the top 12 cities do not change their rank; see text}  \\
\hline
rank& 2007& 2008&2009&2010& 2011\\
   \hline
 13 &    Trieste   &    Trieste   &   Trieste    &   Trieste    &   Parma        \\
14  &    Parma  &    Parma  &   Parma   &   Parma  &   Trieste     \\
15  &    Brescia   &    Brescia   &   Brescia     &   Modena  &   Brescia     \\
16  &    Modena     &    Modena     &   Modena  &   Brescia     &   Modena  \\
17  &    Catania    &    Catania   &   Catania     & Catania     &   Catania     \\
18  &    Messina   & Reggio Emilia  & Messina  & Messina  & Reggio Emilia  \\
19  &    Reggio Emilia  & Messina  & Reggio Emilia  & Reggio Emilia  & Messina  \\
20  &    Prato  &    Prato  &   Prato   &   Prato  & Prato   \\
21  &    Monza  &    Monza  &   Cagliari    &   Perugia   &   Cagliari    \\
22  &    Perugia    &    Perugia    &   Perugia     & Cagliari   &   Perugia     \\
23      &    Cagliari   &    Cagliari   &   Ravenna  & Ravenna     &   Ravenna    \\
24      & Bergamo  & Ravenna   &    Monza   &   Monza  &   Monza    \\
\hline
8080    &    Baradili   &    Falmenta   &   Morterone  &
Falmenta    &   Ribordone  \\
8081    &    Mnt.Leone Rocca Doria  &    Morterone  &
Torresina   &   Morterone    &   Ingria  \\
8082    &    Castelmagno   &    Mnt.Lapiano    &   Canosio
    &   Carapelle Calvisio &   Torresina   \\
8083    &    Torresina &    Canosio    &   Cervatto  &
Cervatto    &   Cervatto   \\
8084    &    Salza Di Pinerolo  &    Elva   &   Salza Di
Pinerolo   &   Cavargna    &   Falmenta    \\
8085    &    Moncenisio     &    Torresina  &   Cavargna
   &   Ingria &   Cavargna    \\
8086    &    Elva   &    Carapelle Calvisio     & Carapelle
Calvisio  &   Torresina   &   Elva   \\
8087    &    Cervatto &    Moncenisio    &   Elva  &
Elva    &   Castelmagno     \\
8088    &    Menarola   &    Cervatto   &   Cursolo-Orasso
&   Cursolo-Orasso  &   Menarola    \\
8089    &    Canosio    &    Menarola   &   Menarola  &
Menarola    &   Cursolo-Orasso  \\
8090    &    Pedesina   &    Pedesina   &   Moncenisio
&   Val Rezzo   &   Val Rezzo   \\
8091    &    Cursolo-Orasso     &    Cursolo-Orasso     &
Pedesina    &   Moncenisio  &   Pedesina   \\
8092    &    Val Rezzo  &    Val Rezzo  &   Val Rezzo  &
Pedesina   &   Moncenisio  \\
\hline \end{tabular} \end{landscape}


\begin{thebibliography}{99}

\bibitem{}
Ausloos, M., 2013.  A scientometrics law about co-authors and their
ranking. The co-author core, Scientometrics 95, 895-909.


\bibitem{}
Baldwin, R., Forslid, R., Martin, P., Ottaviano, G., Nicoud, F.,
2003. Economic geography and public policy. Princeton, NJ: Princeton
University Press.

\bibitem{}
Beckmann, M., 1958. City Hierarchies and the Distribution of City
Size, Economic Development and Cultural Change 6, 243-248.


\bibitem{}
Bosker, M., Brakman, S., Garretsen, H., Schramm, M., 2008. A century
of shocks: the evolution of the German city size distribution
1925-1999, Regional Science and Urban Economics 38(4), 330-347.

\bibitem{}
Brakman, G., Garretsen, H., van Marrewijk, C., van den Berg, M.,
1999. The Return of Zipf: Towards a Further Understanding of the
Rank-Size Distribution, Journal of  Regional Science 39(1), 182-213.

\bibitem{}
Cordoba, J.-C., 2008. On the distribution of city sizes, Journal of
Urban Economics 63(1), 177-197.

\bibitem{}
Cristelli, M., Batty, M., Pietronero, L., 2012. There is more than a
power law in Zipf, Scientific reports 2, 812.

\bibitem{}
Dimitrova, Z., Ausloos, M., 2013. Primacy analysis of the system of
Bulgarian cities, arXiv preprint : $arXiv:1309.0079$.

\bibitem{}
Dobkins, L.H., Ioannides, Y.M., 2001. Spatial interactions among
U.S. cities: 1900-1990, Regional Science and Urban Economics 31(6),
701-731.

\bibitem{}
Eurostat office at: $http://epp.eurostat.ec.europa.eu/$.

\bibitem{}
Fujita, M., Krugman, P., Venables, A.J., 1999. The Spatial
Economics: Cities, Regions, and International Trade, The MIT Press,
Cambridge, MA.

\bibitem{}
Fujita, M., Mori. T., 2005. Frontiers of the New Economic Geography,
Papers in Regional Science 84(3), 377-405.

\bibitem{}
Fujita, M., Thisse, J.-F., 2000. The formation of economic
agglomerations: Old problems and new perspectives, in: Huriot, J.M.,
Thisse, J.-F. (Eds.), Economics of Cities: Theoretical Perspectives,
Cambridge Univ. Press, Cambridge, UK.

\bibitem{}
Gabaix, X., 1999a. Zipf law for Cities: An Explanation, Quarterly
Journal of Economics 114(3), 739-767.

\bibitem{}
Gabaix, X., 1999b. Zipf law and the Growth of Cities American
Economic Review 89(2), 129-132.

\bibitem{}
Gabaix, X., Ioannides, Y.M., 2004.  The Evolution of City Size
Distributions, in: Henderson, J.V., Thisse, J.-F. (Eds.), Handbook
of Regional and Urban Economics, Vol. 4, Amsterdam, Elsevier.

\bibitem{}
Garmestani, A.S., Allen, C.R., Gallagher, C.M., Mittelstaedt, J.D.,
2007. Departures from Gibrat's Law, Discontinuities and City Size
Distributions, Urban Studies 44(10), 1997-2007.

\bibitem{}
Garmestani, A.S., Allen, C.R., Gallagher, C.M., 2008. Power laws,
discontinuities and regional city size distributions, Journal of
Economic Behavior and Organization 68, 209-216.

 \bibitem{}
Gibrat, R., 1931. Les in \'{e}galit\'{e}s  \'{e}conomiques:
Applications aux in\'{e}galit\'{e}s des richesses, \`a  la
concentration des entreprises, aux populations des villes, aux
statistiques des familles, etc., d'une loi nouvelle, la loi de
l'effet proportionnel. Paris: Sirey.

\bibitem{}
Giesen, K., S\"{u}dekum, J., 2011. Zipf law for cities in the
regions and the country, Journal of Economic Geography 11(4),
667-686.

\bibitem{}
Gu\'{e}rin-Pace, F., 1995. Rank-size distribution and the process of
urban growth, Urban Studies 32(3), 551-562.

\bibitem{}
Ioannides, Y.M., Overman, H.G., 2003. Zipf's law for cities: an empirical examination, Regional Science and Urban Economics 33(2),
127-137.

\bibitem{}
Ioannides, Y.M., Skouras, S., 2013. US city size distribution:
robustly Pareto, but only in the tail, Journal of Urban Economics
73(1), 18-29.

\bibitem{}
Italian regions and municipalities at:
$http://www.comuni-italiani.it/regioni.html$

\bibitem{}
Jefferson, M., 1939. The Law of Primate City, Geographical Review 29(2), 226-232.

\bibitem{}
Krugman, P., 1991. Geography and Trade, Cambridge MA: The MIT Press.

\bibitem{}
Krugman, P., 1995. Development, Geography, and Economic Theory,
Cambridge MA: The MIT Press.


\bibitem{}
Laherrere,  J., Sornette, D., 1998.
Stretched exponential distributions in nature and economy
fat tails with characteristic scales, European Physics Journal B 2(4), 525-539.



 \bibitem{ }
Levenberg, K., 1944. A method for the solution of certain problems
in least squares, Quarterly Applied Mathematics 2, 164-168.

 \bibitem{ }
L{\'o}pez-Bazo, E., Vay{\'a}, E., Mora, A.J., Surinach, J., 1999.
Regional economic dynamics and convergence in the European Union,
The Annals of Regional Science 33(3), 343-370.


\bibitem{ }  
Lourakis, M.I.A., 2011.  A Brief Description of the
Levenberg-Marquardt Algorithm Implemented by levmar, Foundation of
Research and Technology 4, 1-6.

  \bibitem{ }
Marquardt, D.W., 1963. An Algorithm for Least-Squares Estimation of
Nonlinear Parameters, Journal of the Society for Industrial and
Applied Mathematics  11(2), 431-441.

\bibitem{}
Matlaba, V. J., Holmes, M. J., McCann, P., Poot, J., 2013. A century
of the evolution of the urban system in Brazil, Review of Urban and
Regional Development Studies 25(3), 129-151.

\bibitem{}
McCann, P., 2013. Modern Urban and Regional Economics, 2nd Edition.
Oxford University Press.

\bibitem{}
Mills, E. S., Hamilton, B.W., 1994. Urban Economics,
Prentice Hall.

\bibitem{}
Neary, J.P., 2001. Of Hype and Hyperbolas: Introducing the New
Economic Geography, Journal of Economic Literature 39(2), 536-561.

\bibitem{}
Nitsch, V., 2005. Zipf zipped, Journal of Urban Economics 57(11),
86-100.

\bibitem{}
Ottaviano, G., Puga, D., 1998. Agglomeration in the global economy:
A survey of the 'new economic geography', The World Economy 21(6),
707-731.

\bibitem{}
Peng, G., 2010. Zipf's law for Chinese cities: Rolling sample
regressions, Physica A: Statistical Mechanics and its Applications 389(18), 3804-3813.

\bibitem{ }
Popescu, I., 2003. On a Zipf's law extension to impact factors.
Glottometrics 6, 83-93.

\bibitem{}
Reed, W.J., 2002. On the Rank-Size Distribution for Human
Settlements, Journal of Regional Science 42(1), 1-17.

\bibitem{}
Rosen, K.T., Resnick, M., 1980.
 The size distribution of cities: an
examination of the Pareto law and primacy, Journal of Urban
Economics 8(2), 165-186.

\bibitem{}
Sheppard, E., 1985. Urban System Population Dynamics: Incorporating
Nonlinearities, Geographical Analysis 17(1), 47-73.

\bibitem{}
Simon, H., 1955. On a Class of Skew Distribution Functions,
Biometrika 42(314), 425-440.

\bibitem{}
Skipper, R.K., 2011. Zipf's  Law and Its Correlation to the GDP of
Nations, McNair Scholars Undergraduate Research Journal 3, 217-226.

\bibitem{}
Song, S., Zhang, K.H., 2002.
 Urbanisation and city size distribution
in China, Urban Studies 39(12), 2317-2327.

\bibitem{}
Soo, K.T., 2007. Zipf law and urban growth in Malaysia, Urban
Studies 44(1), 1-14.


\bibitem{}
Zipf, G.K., 1935. The Psychobiology of Language, Houghton-Mifflin.

\bibitem{}
Zipf, G., 1949. Human Behavior and the Principle of Least Effort,
Cambridge, MA: Addison-Wesley Press.








\end{thebibliography}
\end{document}